\documentclass[aps,prd,superscriptaddress,showpacs,nofootinbib,twocolumn,floatfix,preprintnumbers,]{revtex4-1}

\usepackage{mciteplus}
\usepackage{latexsym}
\usepackage{amsthm}
\usepackage{amsmath}
\usepackage{amssymb}
\usepackage{hepunits}
\usepackage{hyperref}
\usepackage{bbm}
\usepackage{bm}
\usepackage{xfrac}
\usepackage{color}
\usepackage[dvipsnames]{xcolor}
\usepackage{colordvi}
\usepackage{comment}
\usepackage{dcolumn}
\usepackage{times,latexsym,graphicx,wrapfig,url}
\usepackage{epsfig,lineno,bm}
\usepackage{subfigure}
\usepackage{slashed}

\newcommand{\be}{\begin{equation}}
\newcommand{\ee}{\end{equation}}
\newcommand{\bea}{\begin{eqnarray}}
\newcommand{\eea}{\end{eqnarray}}

\newcommand{\bef}{\begin{figure}[htbp]\begin{center}}
\newcommand{\eef}{\end{center}\end{figure}}
\def\lsim{\mathrel{\rlap{\lower4pt\hbox{\hskip1pt$\sim$}}
    \raise1pt\hbox{$<$}}}
\def\gsim{\mathrel{\rlap{\lower4pt\hbox{\hskip1pt$\sim$}}
    \raise1pt\hbox{$>$}}}

\newcommand\UMich{University of Michigan, Ann Arbor, MI }
\newcommand\Princeton{ Princeton University, Princeton, NJ }
\newcommand\FNAL{Fermi National Accelerator Laboratory, Batavia, IL }

\newcommand{\Eq}[1]{Eq.~(\ref{#1})}

\def\lsim{\mathrel{\rlap{\lower4pt\hbox{\hskip 0.5 pt$\sim$}}
\raise1pt\hbox{$<$}}}  

\newcommand{\apr}{A^\prime}

\newcommand{\mzero}{m^0}

\newcommand{\JSNS}{JSNS$^2$}

\begin{document}

\title{Signatures of Pseudo-Dirac Dark Matter at High-Intensity Neutrino Experiments}
\preprint{FERMILAB-PUB-18-148-A, PUPT 2563}

\author{Johnathon R. Jordan}
\email{jrlowery@umich.edu}
\affiliation{\UMich}
\author{Yonatan Kahn}
\email{ykahn@princeton.edu}
\affiliation{\Princeton}
\author{Gordan Krnjaic}
\email{krnjaicg@fnal.gov}
\affiliation{\FNAL}
\author{Matthew Moschella}
\email{moschella@princeton.edu}
\affiliation{\Princeton}
\author{Joshua Spitz}
\email{spitzj@umich.edu}
\affiliation{\UMich}

\begin{abstract}

We (re)consider the sensitivity of past (LSND) and future (JSNS$^2$) beam dump neutrino experiments to two models of MeV-scale pseudo-Dirac dark matter. Both LSND and JSNS$^2$ are close (24-30~m) to intense sources of light neutral mesons which may decay to dark matter via interactions involving a light mediator or dipole operators. The dark matter can then scatter or decay inside of the nearby detector. We show that the higher beam energy of JSNS$^2$ and resulting $\eta$ production can improve on the reach of LSND for light-mediator models with dark matter masses greater than $m_\pi/2$. Further, we find that both existing LSND and future JSNS$^2$ measurements can severely constrain the viable parameter space for a recently-proposed model of dipole dark matter which could explain the 3.5 keV excess reported in observations of stacked galaxy clusters and the Galactic Center.

\end{abstract}

\maketitle


\section{Introduction}
\label{sec:Introduction}

Although there is overwhelming gravitational evidence for the existence of dark matter (DM), its microscopic properties remain elusive despite decades of direct and indirect detection searches (see Ref.~\cite{Bertone:2016nfn} for a historical review). In recent years, beam dump experiments have emerged as powerful probes of dark matter (DM) below the GeV scale, thereby opening up a new frontier in the discovery effort. In these experiments, a beam of protons \cite{Batell:2009di,deNiverville:2011it,Kahn:2014sra,deNiverville:2016rqh,Aguilar-Arevalo:2017mqx} or electrons \cite{Izaguirre:2013uxa,Batell:2014mga,Battaglieri:2016ggd} impinges on a fixed target possibly yielding a secondary beam of DM particles that scatter or decay in a downstream detector -- see Refs.~\cite{Alexander:2016aln,Battaglieri:2017aum} for an overview. Compared to missing energy techniques at fixed target experiments \cite{Izaguirre:2014bca,Gninenko:2016kpg,Banerjee:2016tad} and $B$-factories \cite{Izaguirre:2013uxa,Essig:2013vha,Lees:2017lec}, where the experimental signature is exclusively anomalous energy loss, these ``production-and-detection'' experiments are a more direct probe of DM because they are able to observe the DM directly through scattering or decay signatures in a downstream detector.

\begin{figure*}
\hspace{-1cm}
\includegraphics[width=0.9\textwidth]{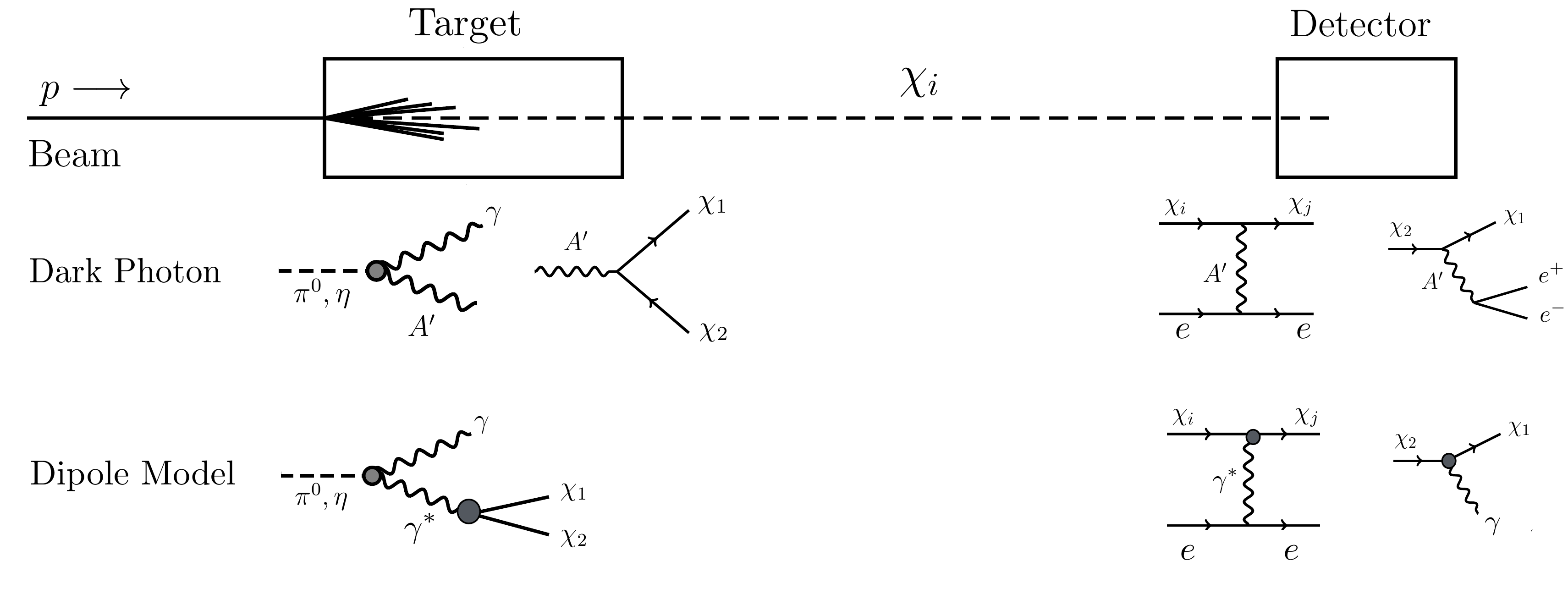}
\caption{Schematic cartoon of the production and detection processes for the dark photon and dipole models described in Sec. \ref{sec:DMProd}. A high energy
proton beam impinges on a fixed target (beam dump) and produces neutral mesons $m^0 = \pi^0, \eta$ which can decay to 
dark sector particles $\chi_1\chi_2$. In the dark photon models this decay is two-step $m^0 \to \gamma A^\prime \to \gamma \chi_1\chi_2$, whereas
for the dipole interaction, the initial meson decay is three-body $m^0 \to \gamma \chi_1\chi_2$ through a virtual photon.  For both 
representative models, the  signal arises from $\chi_{i}$ depositing visible energy inside the downstream detector either as a $\chi_i e \to \chi_j e$ scattering 
process or as a decay, $\chi_2 \rightarrow \chi_1 e^+ e^-$ or $\chi_2 \rightarrow \chi_1 \gamma$. Note that for the dipole model, the $\gamma \chi_1 \chi_2$ interaction is labeled with a gray circle to reflect the fact that this coupling is nonrenormalizable.}
\label{FIG:Cartoon}
\end{figure*}

Unlike traditional direct detection techniques, whose sensitivity is limited by the low momentum transfers imparted by light DM particles traveling at $v \sim 10^{-3} c$ in the 
Galactic Halo, the relativistic kinematics at beam dump experiments make it possible to probe models which would otherwise be undetectable due to the non-relativistic kinematics of cosmological DM. This feature is particularly useful for studying models with predominantly \emph{inelastic} interactions in which the DM couples to the mediator through off-diagonal interactions with a heavier dark-sector particle. At beam dump experiments, there are two principal observables that such models can induce:
\begin{itemize}
\item{\bf Decay Signatures:}
Both light and heavy DM states are generically produced together at the beam dump. If the heavier state is sufficiently long-lived, it can survive to the downstream detector and decay to partially visible final states inside the detector. This is also the strategy of long-lived particle searches at high-energy colliders \cite{Chou:2016lxi,Feng:2017uoz,Gligorov:2017nwh}. The advantage of beam dump experiments is in their extremely high luminosity, which permits detection of decay signatures even when the decay length is much larger than the size of the experiment.
 \item {\bf Scattering Signatures: }
 A sufficiently boosted beam of DM particles has enough energy to inelastically up/down-scatter off Standard Model (SM) particles and deposit copious amounts of visible energy inside the detector. For sufficiently large mass splittings, this process is kinematically forbidden in direct detection experiments, but unsuppressed at beam dumps where the DM is relativistic.
\end{itemize} 
In Fig.~\ref{FIG:Cartoon} we present a schematic cartoon of inelastic DM production and detection  at proton beam dump experiments. The complementarity of
experimental beam dump scattering and decay signatures for various dark matter models was also considered in Refs.~\cite{Morrissey:2014yma,Izaguirre:2017bqb,DAgnolo:2018wcn}.

The 170~ton Liquid Scintillator Neutrino Detector (LSND) experiment~\cite{LSNDDetector}, a fixed-target experiment with a detector situated 30~m from an 800~MeV proton beam, took data from 1993--1998 and currently provides some of the strongest constraints on DM below the $\sim$100 MeV scale~\cite{Auerbach:2001wg}. Such DM can be produced from $\pi^0$ decays at the target and then interact in the detector~\cite{deNiverville:2011it,Izaguirre:2017bqb}. The excellent reach of LSND is primarily due to the large detector mass and high beam power ($\sim3\times10^{22}$~protons on target (POT)/year)~\cite{Aguilar:2001ty}, resulting in significant neutral pion creation ($\sim$0.1 $\pi^0$/POT), but many improvements are possible in future experimental programs. These include higher beam energy to access heavier DM, optimized electron recoil cuts to maximize signal-to-background for various DM masses~\cite{Kahn:2014sra}, and better background rejection from events which fake elastic electron recoils. The J-PARC Sterile Neutrino Search at the J-PARC Spallation Neutron Source (\JSNS) experiment~\cite{Ajimura:2017fld}, which will start data taking with a 3~GeV kinetic energy proton beam in 2019, may achieve some or all of these enhancements to DM sensitivity.

In this paper, we evaluate the reach of \JSNS\ to models of MeV-scale dark matter. To make contact with previous work, we will consider DM which can be produced from light neutral mesons $m^0 = \pi^0, \eta$ ($m_{\pi^0}=134.98$~MeV and $m_{\eta}=547.86$~MeV). We study two representative models: a dark photon model, where mixing between the photon and dark photon $A'$ leads to decay modes $m^0 \to \gamma A' \to \gamma \chi_1 \chi_2$, and a dipole model, where DM interacts directly with the photon through a dimension-5 operator and is produced via $m^0 \to \gamma \gamma^* \to \gamma \chi_1 \chi_2$. To keep the discussion general, we will allow $\chi_1$ and $\chi_2$ to form a pseudo-Dirac pair with arbitrary mass splitting $\Delta = m_2 - m_1$, with the elastic case $m_1 = m_2$ a particular realization of this scenario. We will find that while a higher beam energy allows the production of DM with $m_\pi < m_{1} + m_{2} < m_\eta$ through $\eta$ decays (a mode inaccessible to LSND, which operated below $\eta$ production threshold), the additional neutrino backgrounds from mesons that do \emph{not} produce DM from rare decays (e.g. kaons, also not produced significantly at LSND) tend to degrade the reach for light DM at lower masses. However, a medium-energy experiment like \JSNS\ serves an important role in covering parameter space inaccessible to both LSND and the higher-energy (8~GeV beam) MiniBooNE experiment~\cite{Aguilar-Arevalo:2017mqx}.

The dark photon model has been well studied in multiple scenarios~\cite{Alexander:2016aln,Battaglieri:2017aum}, and the dipole model has recently attracted attention as a possible explanation for the excess of 3.5 keV gamma rays from the Galactic Center and the Perseus Cluster~\cite{DEramo:2016gqz}. While it should be noted that UV completions of the dipole model have already been strongly constrained by collider experiments~\cite{Izaguirre:2015zva}, beam dump experiments can test this model directly as the operator that sources the 3.5 keV line also enables DM production from meson decays and scattering with detector electrons. Re-evaluating the LSND data in light of this model, we will show that LSND already rules out large parts of the preferred parameter space, with \JSNS\ able to cover a similar region in the near future.

This paper is organized as follows. In Sec.~\ref{sec:DMProd}, we discuss the representative DM models along with the production mechanisms and detection signals from proton beam dumps. In Sec.~\ref{sec:JSNS}, we describe the \JSNS\ experimental setup, including the beam dump and neutrino detector. In Sec.~\ref{sec:Backgrounds}, we describe the backgrounds to a DM search at \JSNS, consisting primarily of neutrinos produced in the target and cosmic rays. In Sec.~\ref{sec:Reach}, we present the projected reach of \JSNS\ to the representative DM models, and compare with previous results and a new reanalysis of LSND data within the dipole DM model. We conclude in Sec.~\ref{sec:Conc}. Further details of the matrix elements used in our reach projections are given in Appendix~\ref{app:Matrix}.




\section{DM production and detection}

\label{sec:DMProd}
\subsection{Representative pseudo-Dirac models}
We suppose the DM components of our model consist of mass eigenstates $\chi_1$ and $\chi_2$, with masses $m_1$ and $m_2$, respectively, and mass splitting $\Delta = m_2 - m_1$. 
Such a mass splitting naturally arises for fermionic fields with both Dirac and Majorana masses. For instance, a Dirac spinor with $\psi = (\xi, \eta^\dagger)$ built out of 
two Weyl spinors $\xi$ and $\eta$ can have the following mass terms in the interaction basis:
\bea
-{\cal L}_{\rm mass} = m \xi \eta  + \frac{\mu_\xi }{2}     \xi \xi +   \frac{\mu_\eta}{2} \eta\eta + h.c. ,
\eea
where $m$ is the Dirac mass and  $\mu_i$  is the Majorana mass for each component.
In the $\mu_\xi = \mu_\eta  \equiv \mu$ limit, the mass eigenstates for this system are
\bea 
\chi_1 = \frac{i}{\sqrt{2}}(\eta - \xi)~~~~,~~~~\chi_2 = \frac{1}{\sqrt{2}} ( \eta + \xi ) ~,~
\eea
with corresponding eigenvalues 
$m_{1,2} = m \mp \mu$.

If the  interaction-basis spinor $\psi$ couples to other particles through dark currents of the form $ \overline \psi \Gamma \psi$, where $\Gamma = \{\gamma^\mu, \sigma^{\mu\nu}, i \sigma^{\mu\nu} \gamma^5\}$, 
 then to leading order in $\Delta/m$, these interactions will  naturally be off-diagonal in the mass basis
 \bea
\overline \psi \Gamma \psi   ~\to~
 \overline \chi_2 \Gamma \chi_1  + h.c.,
 \eea
which yields inelastic scattering processes that interconvert $\chi_{1,2}$ and induces $\chi_2$ decays if 
the latter are kinematically allowed. For the remainder of this paper we will couple dark currents of this form to SM fields.
 
 We consider two representative scenarios: 
\begin{itemize}
\item \textbf{Dark photon.} The DM has an off-diagonal coupling to a new U(1) gauge boson $A'$,
\be
\mathcal{L} \supset g_D A^\prime_\mu \overline{\chi}_2\gamma^\mu \chi_1 + {\rm h.c.} ~,
\ee
while the $A'$ interacts with the SM through the EM current,
\be
\mathcal{L} \supset \epsilon e A'_\mu J^\mu_{\rm EM}.
\ee
In particular, this means that an $A'$ may replace a photon in any SM process, at the cost of a factor of the kinetic mixing parameter $\epsilon$. For more details about this model, including a UV completion, see for example Ref.~\cite{Izaguirre:2017bqb}. Throughout our analysis, we will assume that the new U(1) is spontaneously broken and that the dark photon has a mass $m_{A^\prime} > m_1 + m_2$ so that $A^\prime \to \chi_1 \chi_2$ decays 
are kinematically allowed. 
\item \textbf{Dipole interaction.} Here, the DM couples directly to the photon through dimension-5 electric or magnetic dipole operators,
\be
\mathcal{L} \supset -\frac{i}{2\Lambda} \overline{\chi}_2 \sigma^{\mu \nu}\left(c_M + i c_E \gamma^5\right)\chi_1 F_{\mu \nu},
\label{eq:DipoleOp}
\ee
where $\sigma^{\mu \nu} = \frac{i}{2}[\gamma^\mu, \gamma^\nu]$. DM with electric and/or magnetic dipole moments was first considered in Ref.~\cite{Sigurdson:2004zp}, and collisionally-excited DM was considered in Ref.~\cite{Finkbeiner:2007kk}. This dipole DM model considered in Ref.~\cite{DEramo:2016gqz}, with $\Delta = 3.5 \ \keV$, combines these features such that collisional excitation of $\chi_1$ to $\chi_2$ followed by the decay $\chi_2 \to \chi_1  \gamma$ would explain the 3.5 keV line.
\end{itemize}
In both scenarios, the relic density of stable $\chi_1$ particles can arise from $\chi_1 \chi_2 \to \overline f f$ coannihilation to charged SM fermions $f$. 
The Boltzmann equation governing freeze-out for each species is 
\bea
\label{eq:boltzmann}
\frac{dn_{i}}{dt} + 3H n_{i} &=& - \langle \sigma v\rangle_{\rm coann.} \left(  n_1 n_2 - n_1^{\rm (eq)}  n_2^{\rm (eq)}       \right) \nonumber \\ 
&& \hspace{-1.35cm}\pm \bigl(    n_f  \langle \sigma v \rangle_{\rm scat.}     + \langle \Gamma_{\chi_{2}} \rangle   \bigr) \left( n_2 - n_2^{\rm (eq)} \right) ,~~~~~~     
\eea
where $n_i$ is the number density for species $i$, $\langle \sigma v\rangle_{\rm coann}$ is the 
 $\chi_1 \chi_2 \to \overline f f$ coannihilation cross section, $\langle \sigma v\rangle_{\rm scat}$ is the $\chi_2 f \to \chi_1 f$ scattering cross section, 
 and $\langle \Gamma_{\chi_{2} }\rangle$ is the $\chi_2 \to \chi_1 + {\rm SM}$ decay rate. In this equation, an $^{(\rm eq)}$ superscript denotes an equilibrium quantity, $\langle \cdots  \rangle$ brackets represent thermal averages, 
 and the upper/lower signs correspond to $i=1,2$ respectively. 
 Solving this system for the dark photon model~\cite{Izaguirre:2017bqb} and the dipole model~\cite{DEramo:2016gqz}
 yields predictive relic density targets accessible to beam dump experiments. 

\subsection{Beam dump production: meson decay}
The coupling of DM to the photon (either directly through dipole interactions, or indirectly through the $A'$) allows the decays
\be
\pi^0, \eta \to \gamma \chi_1 \chi_2.
\ee
Feynman diagrams for these processes are shown in Fig.~\ref{FIG:Cartoon}. Note that in the dipole model, the decay proceeds through a virtual $\gamma^*$, while in the dark photon model, the $A'$ can either be on- or off-shell depending on its mass. We generate DM events by using the Li\`{e}ge Intranuclear Cascade model (\texttt{INCL})~\cite{Mancusi:2014} to model interactions of the \JSNS\ proton beam with a mercury target, and decay the resultant mesons using the matrix elements given in App.~\ref{app:Matrix}.

We note that there is an additional production mode from dark or ordinary bremsstrahlung in the target, $p N \to p N (A',\gamma) \to p N \chi_1 \chi_2$. However, the 3 GeV beam energy of \JSNS\ makes a reliable simulation difficult \cite{deNiverville:2016rqh}. Therefore we only consider DM production from meson decay, and our DM sensitivity results can be considered conservative and may improve at higher DM masses with the addition of this production mode.
\subsection{Signals: scattering and decay}
\label{sec:signals}
There are several possible DM-related signals for a generic beam dump neutrino experiment depending on the capabilities of the detector and the mass splitting $\Delta$. These are discussed below, assuming $E_{\rm beam} = 3 \ \GeV$ as at \JSNS. Based on the capabilities of \JSNS, the main signal criterion is the total visible energy $E_e$ from all electrons and positrons in the final state; see Sec.~\ref{sec:JSNS} for more details on detector performance.
\begin{itemize}
\item $\bm{\Delta < 2m_e}$. Since $\Delta \ll E_{\rm beam}$, $\chi_{1,2}$ are highly boosted and the mass splitting does not appreciably affect the kinematics. The principal interactions in the detector are (quasi)-elastic scattering, $\chi e \to \chi e$. This notation is meant to indicate that $\chi_1 e \to \chi_2 e$ and $\chi_2 e \to \chi_1 e$ are functionally equivalent, so we do not distinguish between them. In the dipole model, if $\chi_2$ survives to the detector, a decay inside the detector $\chi_2 \to \chi_1 \gamma$ may result in a photon-induced signal, which can be distinguishable from elastic scattering if the detector, unlike \JSNS, has the particle identification capability to separate $\gamma$ and $e$.\footnote{The lowest-order decay process in the dark photon model is $\chi_2 \to \chi_1 + 3\gamma$, with a lifetime exceeding the age of the universe for the relevant parameter space.} However, because of the small mass splitting, the rate for this decay is small enough as to negligibly affect the projected reach, so we do not consider it further.
\item $\bm{\Delta > 2m_e}$. As shown in Ref.~\cite{Izaguirre:2017bqb}, for an LSND-like experiment (including JSNS$^2$), the dominant signal is $\chi_2$ decay inside the detector, $\chi_2 \to \chi_1 e^+ e^-$. For a detector without angular reconstruction capabilities like \JSNS, the electron and positron signals cannot be separated and this mode will look just like elastic electron scattering. However, the signal strength is independent of the target density inside the detector, and depends only on the geometry of the detector with respect to the beam dump. For sufficiently large values of $\epsilon$, all $\chi_2$ will have decayed before reaching the detector, and up-scattering $\chi_1 e \to \chi_2 e$ is kinematically forbidden for large $\Delta$, so for some $m_1$ and $\Delta$ there may be a maximum value of $\epsilon$ which can be probed by the experiment.

Several other signals are possible, including upscattering $\chi_1 N \to \chi_2 N$ where $N$ is a nucleon or a nucleus, followed by $\chi_2$ decay, but the presence of e.g. an additional proton in the final state may cause the event to be classified as background (see Sec.~\ref{sec:Backgrounds}). Such scattering channels would require a dedicated analysis, so for this paper, we focus exclusively on final states with only electrons and/or positrons.
\end{itemize}




\section{The \JSNS\ experiment}
\label{sec:JSNS}

 \JSNS\ is a next-generation short baseline neutrino oscillation experiment designed to search for high-$\Delta m^2$ oscillations ($\sim 0.1-100$ eV$^2$) and measure neutrino cross sections~\cite{Ajimura:2017fld}. \JSNS\ will search for $\overline{\nu}_e$ appearance using neutrinos from muon decay at rest ($\mu^+ \rightarrow e^+ \nu_e \overline{\nu}_\mu;~\overline{\nu}_\mu \rightarrow \overline{\nu}_e$), similar to LSND. In the initial phase, a single detector will be placed on the third floor of the J-PARC Material and Life Science Experimental Facility (MLF), 24 m from the production target which also acts as the beam dump. In addition to providing the neutrinos for the primary physics goals of \JSNS\ (see Ref.~\cite{Ajimura:2017fld}), the MLF beamline is an intense source of light neutral mesons ($\pi^0, \eta$) which can be used to test the DM models outlined here.

\subsection{Beam dump}

The  J-PARC MLF features a 3 GeV proton beam from the Rapid Cycling Sychrotron (RCS) incident on a mercury spallation neutron target. After significant upgrades to the target in the near future, the beam is expected to reach an intensity of 1~MW corresponding to $3.8 \times 10^{22}$ POT/year assuming 5000 hours/year of operation. As of June 2018, the MLF beam power is 525~kW. Protons are produced with a repetition rate of 25 Hz and each beam spill contains two 100 ns pulses separated by 540 ns~\cite{Ajimura:2017fld}. This timing structure provides a very low duty factor, defined as the ratio of beam-on to beam-off ($5\times 10^{-6}$), especially when compared to previous experiments like LSND ($0.07$)~\cite{Athanassopoulos:1997pv}. A low duty factor enables a significant reduction in steady-state, beam-off backgrounds such as cosmic rays. Further, the tight beam pulse windows allow for discrimination between ``on-bunch" activity coming from both beam-based non-neutrino and pion and kaon decay induced neutrinos and ``off-bunch" activity coming from the (longer-lived, $\tau=2.2~\mu$s) muon decays. The beginning of the beam pulse windows are so tight that relativistic DM will often arrive before pion decay-at-rest neutrinos, given the $\pi^+$ lifetime ($\tau=26$~ns), providing another possible analysis tool for mitigating neutrino-induced background.

Due to the high intensity and energy of the beam, the MLF produces large quantities of both $\pi^0$ and $\eta$ mesons. By comparison, the total number of $\pi^0$ produced over the lifetime of the LSND experiment is less than the integrated $\pi^0$ luminosity over one year at the MLF, and no $\eta$ production was expected at LSND.
A simulation of 3~GeV protons incident on mercury using \texttt{INCL++} predicts $0.585\ \pi^{0}$/POT and $0.035\ \eta$/POT corresponding to $2.2 \times 10^{22}\ \pi^0$/year and $1.3 \times 10^{21}\ \eta$/year. Toy simulations using \texttt{Geant4}~\cite{Agostinelli:2003} predict similar $\pi^0$ yields with small fluctuations depending on the choice of physics list, but $\eta$ production is only modeled correctly by \texttt{INCL}. For the signal estimates discussed below, we use the standalone \texttt{INCL++} package as the current \texttt{Geant4} physics lists incorporating this code are considered experimental.

\begin{figure}[t]
\centering
\includegraphics[width=0.45\textwidth]{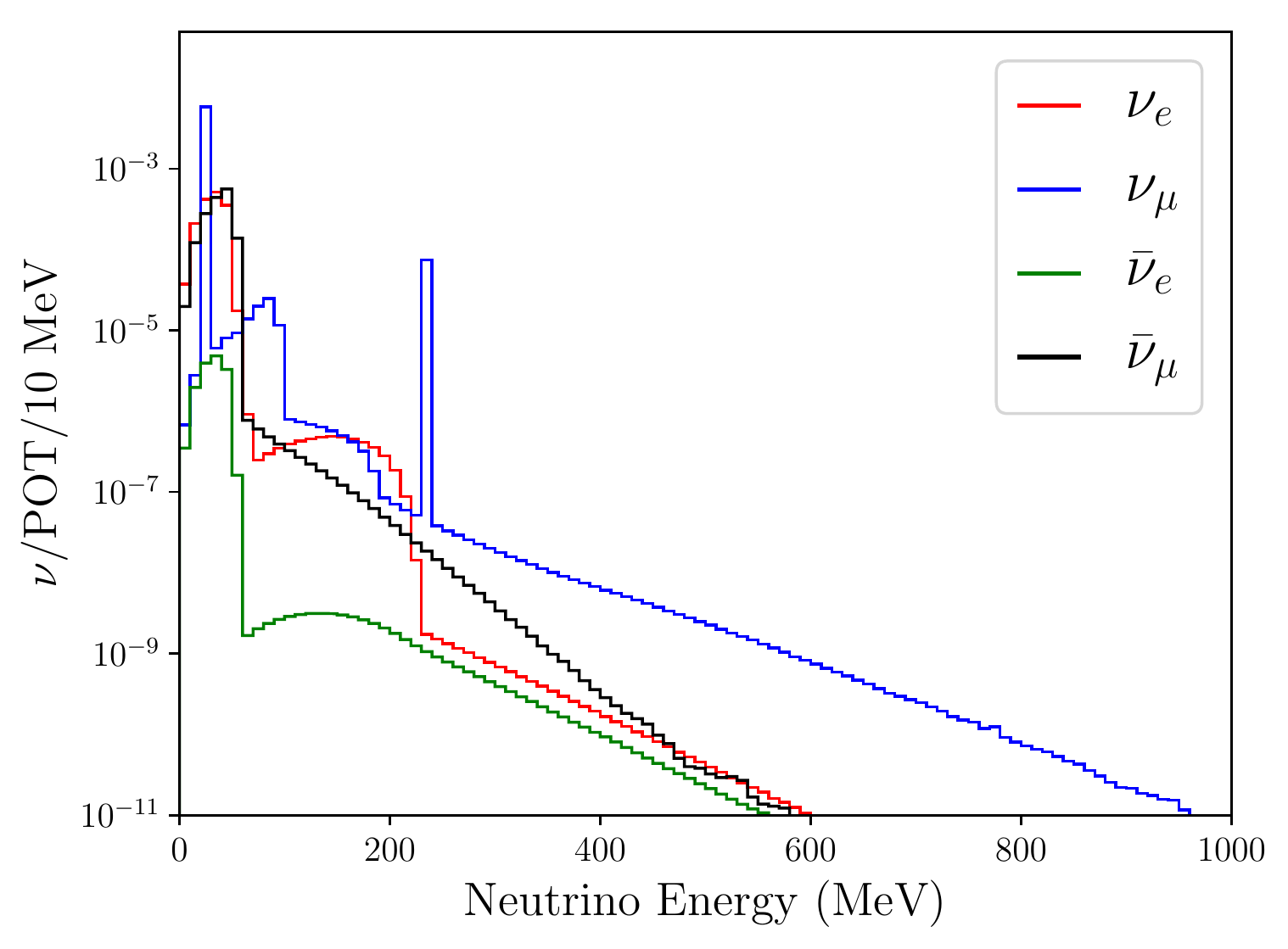}
\caption{The estimated neutrino flux in JSNS$^2$ from the MLF target. The high energy flux is dominated by contributions from pion and kaon decay in flight. The $\nu_\mu$ spike at 236~MeV is due to charged kaon decay at rest [$K^+ \rightarrow \mu^+ \nu_\mu$ (BR=63.6\%)].}
\label{FIG:NeutrinoFlux}
\end{figure} 
\vspace{-0.4cm}

\subsection{Neutrino detector}

\JSNS\ will utilize a liquid scintillator-based detector located 24 m from the mercury target. The detector consists of 3 volumes: an inner acrylic vessel filled with 17 tons of Gd-doped liquid scintillator, a buffer region immediately outside the inner volume, and an outer veto which is optically separated from the buffer and inner acrylic using reflective and absorptive materials. Both the buffer and veto are filled with undoped liquid scintillator totaling approximately 30~tons. 193~8-inch PMTs view the inner volume and 48~5-inch PMTs sit in the veto region to reject activity coming from outside the detector~\cite{Ajimura:2017fld}.

The liquid scintillator used in \JSNS\ will be the same mixture used in the Daya Bay reactor neutrino experiment~\cite{Beriguete:2014}. One of the important features of this specific scintillator is its pulse shape discrimination capability, which allows light particles (electrons and muons) to be distinguished from heavier particles (nucleons)~\cite{Xiao-Bo:2011}. Due to the high light yield of this scintillator ($\sim$10,000 $\gamma$/MeV) and fast time constants (the fast scintillation component has a time constant of $\sim$3.9 ns), it is difficult to isolate the \v{C}erenkov light produced by particles over threshold. As a result, any angular reconstruction, which would normally rely on \v{C}erenkov light, is quite difficult, so we assume no angular reconstruction capabilities, though we note that such capabilities could help distinguish signal from background for the DM searches outlined.

Despite the presence of added lead shielding (10~cm) underneath the detector volume, in addition to the $\sim$20~m of concrete and iron between target and detector, beam-induced and environmental gamma ray backgrounds are expected to be significant below 2.6 MeV. In order to mitigate cosmic gamma ray backgrounds (discussed later in Sec.~\ref{sec:Backgrounds}) and completely remove beam-induced gammas, we consider a detector with an additional 7 cm of lead shielding surrounding the outer stainless steel tank, similar to the design of LSND. This additional shielding is not part of the current \JSNS\ design, but is necessary to sufficiently attenuate the cosmic gamma ray background for the DM searches outlined here. Further, beam-induced fast neutron backgrounds are expected to be significant at low energies based on \textit{in situ} background measurements and detector simulations~\cite{Harada:2016rou}. As a result, we consider a conservative minimum signal energy threshold of 20~MeV which, in combination with the detector's significant active and passive shielding and neutron identification capabilities, render this background negligible. However, a lower threshold, which could substantially enhance sensitivity to DM signals, may be possible when the experiment begins running and mature background rejection techniques are developed.

For visible energies below $\sim$60 MeV, the energy resolution of the detector is given by $\Delta E/E \sim\sqrt{p_0 ^2/E + p_1^2}$ where $p_0$ = 0.07 MeV$^{1/2}$ is a contribution from the number of photoelectrons and $p_1$ = 0.02 is a constant term governed by the detector hardware~\cite{Ajimura:2017fld}. For energies well above the muon decay-at-rest endpoint (52.8~MeV), the energy resolution has not been characterized with detailed measurements. A proper evaluation of the energy resolution at energies above 100 MeV, for example, would need to include PMT and electronics saturation and position dependence, but estimating these effects is beyond the scope of this work. In a simple attempt to account for these effects, we assume that the energy resolution is given by $\sqrt{p_0^2/E + p_1^2 + p_2^2 E}$ where $p_2$ = 0.002 MeV$^{-1/2}$. The added term becomes relevant at high energies and models the degradation of the energy resolution due to saturation. We note that the search proposed here is only weakly dependent on the energy resolution at high energies and a more detailed treatment would not affect the results considerably.


\section{Backgrounds}
\label{sec:Backgrounds}


\begin{figure}
\centering
\includegraphics[width=0.45\textwidth]{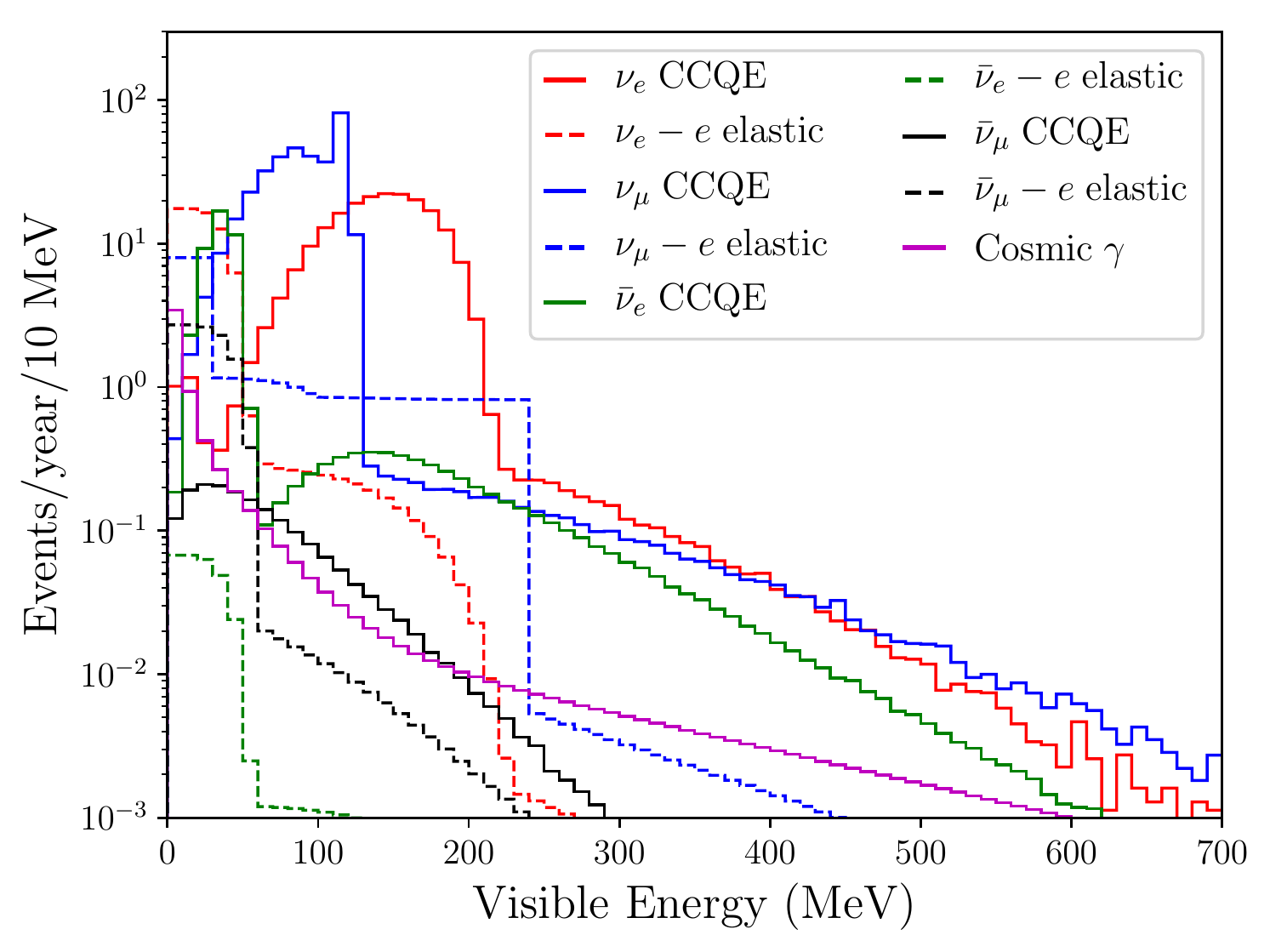}
\caption{Visible energy spectra for the relevant sources of background to the DM signal at JSNS$^2$. A Michel electron cut has been applied to the $\nu_\mu$, $\overline{\nu}_\mu$ CCQE backgrounds and a pulse shape discrimination cut has been applied to the $\nu_\mu$, $\nu_e$ CCQE backgrounds. We assume 7~cm of lead shielding exists around the detector to attenuate the cosmic gamma background.}
\label{FIG:BackgroundRates}
\end{figure} 
\vspace{-0.3cm}

There are two main classes of backgrounds to the DM searches in \JSNS\ proposed in Sec.~\ref{sec:DMProd}: beam-on backgrounds including beam-induced gamma rays and neutrons and neutrinos produced in the target, and steady-state backgrounds coming from cosmic rays and environmental gamma rays. Detailed measurements of many of these backgrounds have been performed using a 500~kg plastic scintillator detector placed on the third floor of the MLF~\cite{Ajimura:2015}. Above the 20~MeV threshold used for this analysis, the dominant backgrounds come from neutrino interactions in the detector and cosmic gamma rays which can fake the DM-induced electron scattering or decay signals.

To estimate the neutrino backgrounds, a \texttt{Geant4}~\cite{Agostinelli:2003} simulation was performed using a detailed MLF target geometry. The flux of all relevant neutrino flavors is shown in Fig.~\ref{FIG:NeutrinoFlux}. Both charged-current  quasi-elastic (CCQE; $\nu_l n \rightarrow l p$ or $\bar{\nu}_l p \rightarrow l n$, where $l$ is either a muon or electron) and neutrino-electron elastic scattering ($\nu e \rightarrow \nu e$) processes are considered. We ignore both resonant and deep inelastic contributions for simplicity, the inclusion of which would not substantially affect our sensitivity estimates. The CCQE cross sections and kinematics were obtained from the NuWro neutrino event generator~\cite{Golan:2012, Juszczak:2009qa}. Events with a muon in the final state usually produce a visible muon-decay-induced Michel electron ($\mu \rightarrow e \nu \nu$). Such events can be rejected efficiently. Requiring the absense of a Michel electron in event selection yields a rejection factor of 15 for $\nu_\mu$ CCQE events, noting that about 6.7\% of $\mu^-$ will capture on nuclei before decaying~\cite{Ajimura:2017fld}. Similarly, a conservative rejection factor of 20 was applied to $\overline{\nu}_\mu$ CCQE events given that they will not suffer from losses due to muon capture.


\begin{figure*}
\centering
\hspace{-0.5cm}\includegraphics[width=0.425\textwidth]{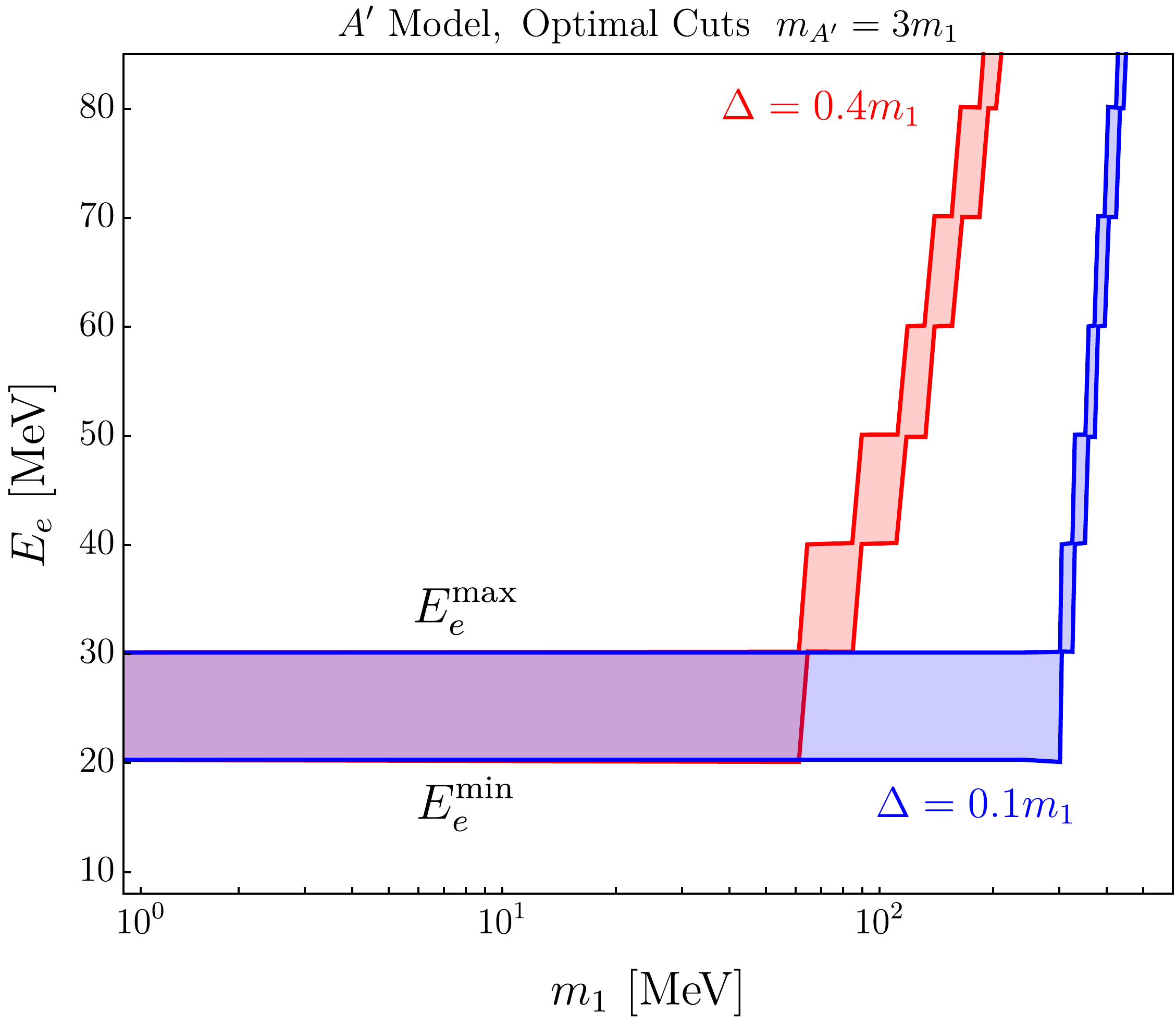}
\hspace{1cm}
\includegraphics[width=0.43\textwidth]{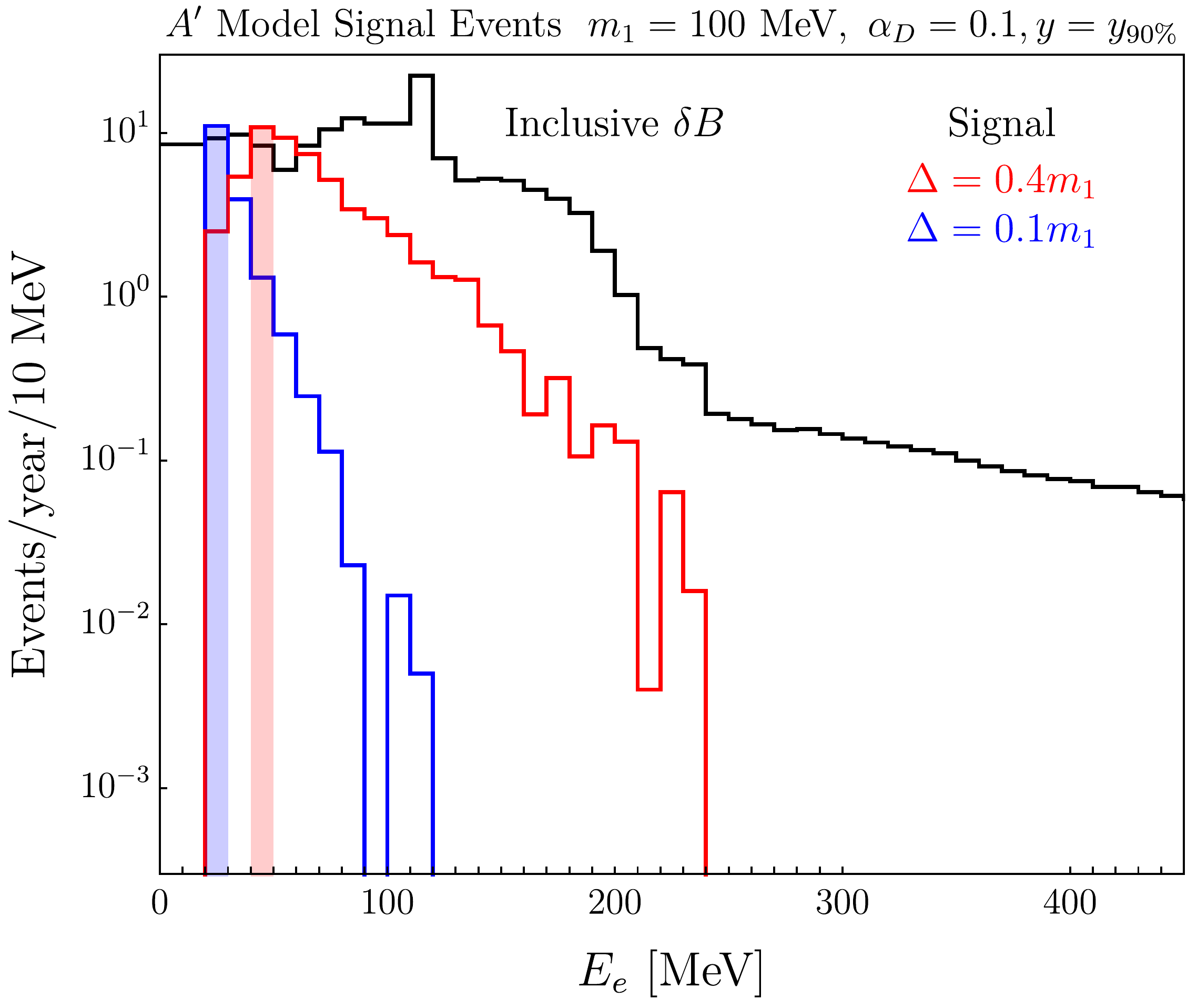}
\vspace{-0.3cm}
\caption{{\bf Left: }   Optimal signal windows in the dark photon model for $\Delta = 0.1 m_1$ (blue shaded region) and $\Delta = 0.4 m_1$ (red shaded region) as a function of $m_1$ for the proposed \JSNS\ search. The dominant processes for visible energy deposit are electron scattering at low masses, and $\chi_2$ decay at high masses.  Here $E^{\rm min}_{e}$ and $E^{\rm max}_{e}$ define the cut interval that maximizes $S/\delta B$ as defined in Eq.~(\ref{eq:detab}). 
{\bf Right:  }     
The electron recoil energy spectrum expected at \JSNS\ for various choices of model parameters. The background uncertainty estimate $\delta B$ is shown for reference. The peak of the signal spectrum typically occurs at $E_e\sim \Delta$ and falls off much more rapidly than the backgrounds for larger energies.
The signal normalization here depends on our choice of $m_1 = 100$ MeV, $\alpha_D = 0.1$, and $y = \epsilon^2 \alpha_D (m_1/m_{A^\prime})^4 = y_{90 \%}$, which 
corresponds to the $90\%$ exclusion contour in Fig.~\ref{fig:DarkPhotonReachDelta} at this mass point. Note that for $\Delta = 0.4$, there are two such values of $y$ for $m_1 = 100 \ \MeV$, but the spectrum is identical for both. The shaded blue and red bins correspond to the optimal cut interval presented on the left panel. }
\label{fig:OptimalCuts}
\end{figure*}

In addition to the simple Michel electron cut, we also apply a pulse shape discrimination cut to the $\nu_\mu$ and $\nu_e$ CCQE backgrounds. These CCQE events will feature both a final state lepton and proton in contrast to signal events which only feature an electron. The proton will produce more delayed scintillation light, allowing CCQE events with energetic protons to be distinguished from signal events. Using the time constants measured in Ref.~\cite{Harada:2016vlb}, we constructed a toy simulation and found that the proton must carry at least 31\% of the total event energy in order for the waveform to be distinguished from a signal event with 90\% purity. When this cut is applied to events generated by NuWro, it results in an additional rejection factor of 2.2 (1.5) for $\nu_\mu$ ($\nu_e$) CCQE events.


\begin{figure*}[t!]
\centering
\includegraphics[width=0.475\textwidth]{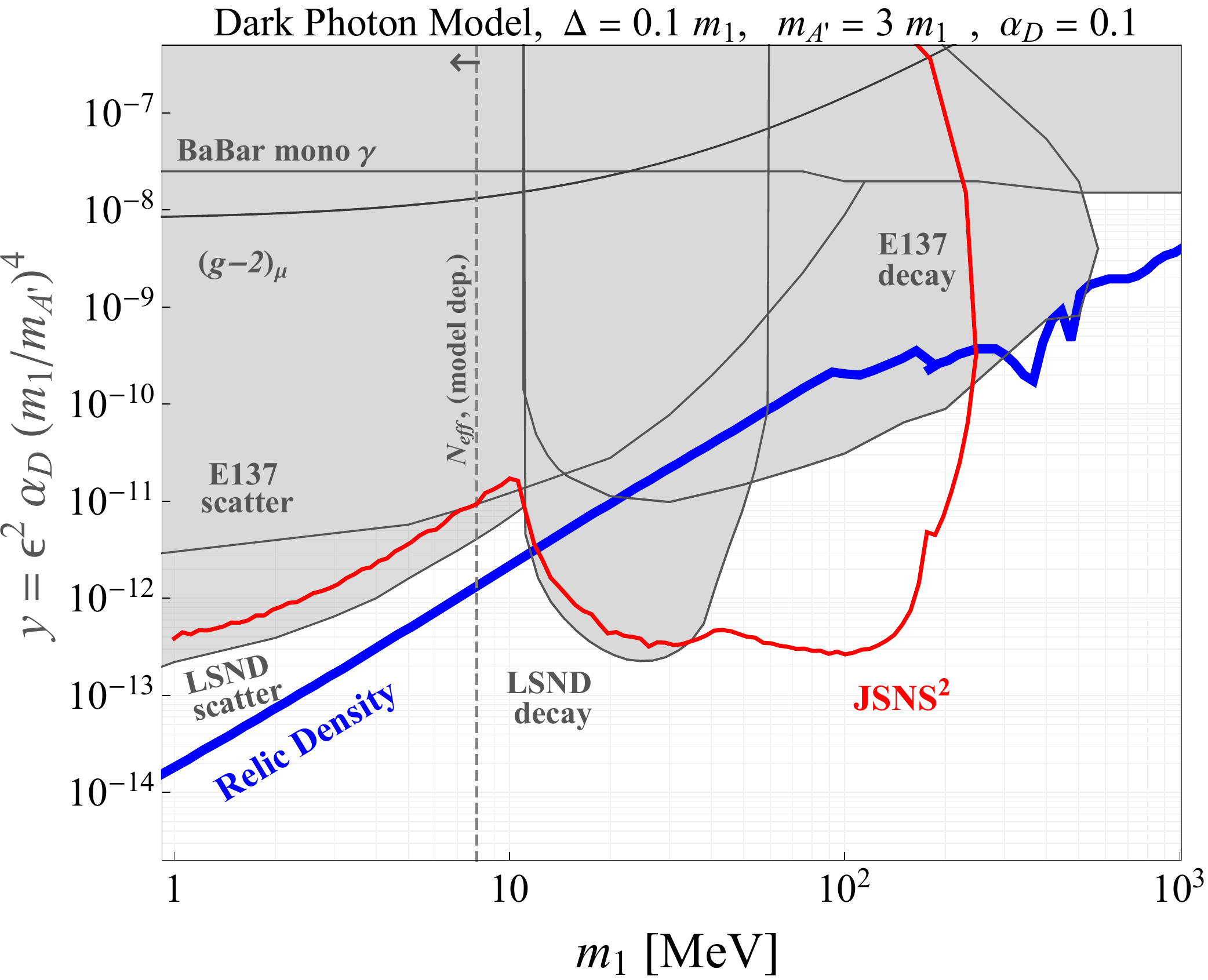}~~~
\includegraphics[width=0.475\textwidth]{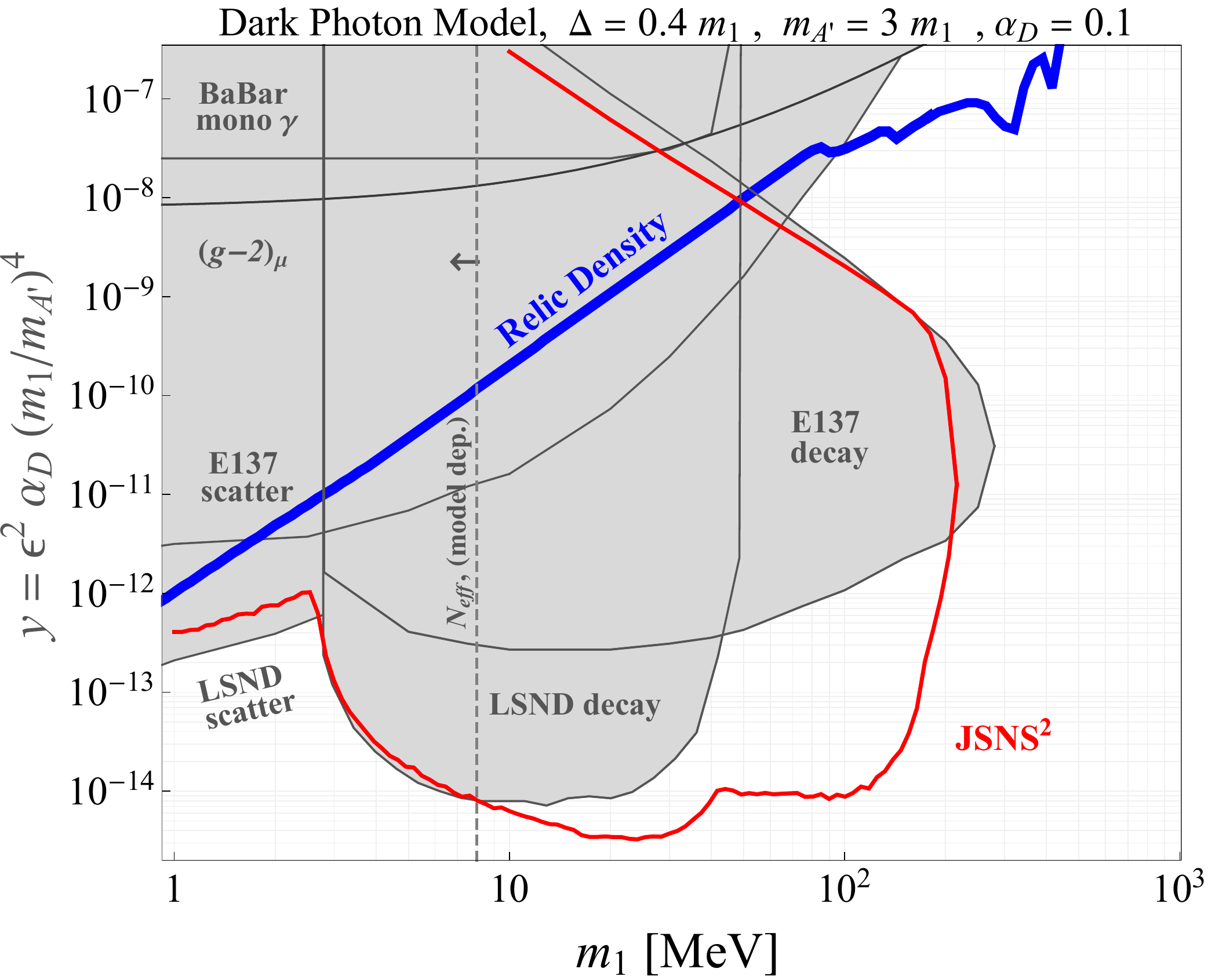}
\caption{Parameter space for thermal pseudo-Dirac DM in the dark photon model for two choices of mass splitting $\Delta$ in terms of the dimensionless parameter $y = \epsilon^2 \alpha_D (m_1/m_{A^\prime})^4$, 
which is proportional to the $\chi_1 \chi_2 \to \bar f f$ annihilation rate in the early universe. The blue curves, computed in Ref.~\cite{Izaguirre:2017bqb}, represent the parameter space for which coannihilation achieves the observed DM relic density. 
The red curves represent the \JSNS\ 1-year reach computed in this paper; other constraints from E137, BaBar, and LSND are shown in gray and taken from Ref.~\cite{Izaguirre:2017bqb}. The red contour for $\Delta = 0.1 m_1$ has the same qualitative behavior as the corresponding region for $\Delta = 0.4 m_1$, only shifted upwards towards $y \sim 10^{-6}$. Note that for both panels, \JSNS\ is superior to the similar LSND experiment in the mass range where $\eta$ decay dominates and $\pi^0$ decay is kinematically forbidden. }
\label{fig:DarkPhotonReachDelta}
\end{figure*}

Steady-state and neutrino-from-beam-muon backgrounds are significantly reduced by windowing the search around the beam pulses. For this study, we consider a 1 $\mu$s window starting at the beginning of the first beam pulse which gives a steady-state rejection factor of 40,000 based on timing alone. However, we note that the neutrinos from muon decay-at-rest, one of the key backgrounds for the search outlined here, could be reduced significantly with an even tighter cut. For example, a 200~ns window after the start of each of the two beam pulses (100~ns each, separated by 540~ns at 25~Hz) would reduce the from-muon neutrino background by a factor of 2.8 with a negligible loss in signal for relativistic DM. The veto is expected to be able to reject cosmic ray muons with a rejection factor of 100~\cite{Ajimura:2017fld} and further rejection power can be achieved by looking for the double coincidence from muon decay. Cosmic neutron backgrounds can be identified and vetoed with a similar rejection factor using pulse shape discrimination, as discussed above. Thus, the only non-negligible cosmic background comes from gamma rays which can pass through the veto without interacting and fake the DM-induced electron scattering or decay signal.

We have studied the cosmic gamma background in detail using the CRY generator~\cite{Hagmann:2007}. A toy simulation of the \JSNS\ veto was used to determine the fraction of cosmic gamma rays which would make it through the veto without interacting. For each cosmic ray generated by CRY, the path length through the veto was determined and the interaction probability was calculated assuming a conversion length of 50~cm in the liquid scintillator. About 60\% of cosmic ray gammas will interact in the veto region, giving a rejection factor of 2.5. We note that it is possible for gammas to make it through the veto region and interact in the buffer region outside the inner volume. It will still be possible to reconstruct some of these events as being outside the fiducial volume so our estimate of the gamma ray background is conservative. Due to the coarseness of the energy bins available to generate cosmic rays in CRY, a triple exponential fit to the resulting cosmic ray spectrum was used. This follows the treatment used to model the measured cosmic ray gamma background below 100 MeV on the MLF third floor~\cite{Ajimura:2015}, but adds an additional exponential to account for the rate at high energies. Finally, due to the extremely high cosmic background event rate even after the veto and timing cuts, we assume that an additional 7~cm of lead shielding is added around the detector to further attenuate the cosmic gamma ray background. In addition, this extra shielding makes the beam-based gamma background negligible.

The final background rates from all sources after cuts can be seen in Fig.~\ref{FIG:BackgroundRates}. We see that with all cuts applied, the cosmic backgrounds are subdominant over the entire visible energy range, and elastic neutrino events dominate at low energies while CCQE neutrino events dominate at high energies. The dominant sources of the high-energy $\nu_e$ and $\nu_\mu$ are kaons, in contrast to the case of the 800 MeV beam considered in Ref.~\cite{Kahn:2014sra} where $\nu_e$ from helicity-suppressed pion decay ($\pi^+ \rightarrow e^+ \nu_e$) dominated at high energies.




\section{Experimental reach}
\label{sec:Reach}
To determine the sensitivity of \JSNS\ to DM, we compute the expected number of signal ($S$) and background ($B$) events over 1 year of running, noting that \JSNS\ is expected to run for 3~years or more. Following the analysis of Ref.~\cite{Kahn:2014sra}, we define the background uncertainty as
\be
\label{eq:detab}
\delta B = \sqrt{B_{\rm beam-off}} + 0.2 B_{\rm beam-on},
\ee
where $B_{\rm beam-off} \approx B_{\rm cosmic}$ is the number of cosmic events faking electron recoils, and $B_{\rm beam-on}$ are all other backgrounds discussed in Sec.~\ref{sec:Backgrounds}, including CCQE and elastic recoil events and accounting for all timing and pulse-shape cuts. Note that the low duty factor $\kappa = t_{\rm on}/t_{\rm off} \ll 1$ for \JSNS\ means the uncertainty on beam-unrelated events is $\delta B_{\rm cosmic} = \sqrt{(1+\kappa)B_{\rm cosmic}} \approx \sqrt{B_{\rm cosmic}}$; this is smaller than that considered in Ref.~\cite{Kahn:2014sra} which had a larger beam timing window. Similarly, to be conservative, we take an overall 20\% systematic uncertainty on all neutrino-related backgrounds, where the CCQE uncertainty arises from nuclear matrix element uncertainties and the elastic uncertainty reflects the uncertainty in the total beam flux.  We define our reach curves by $S/\delta B = 1.3$.\footnote{This roughly corresponds to the 90\% confidence level limit of the LSND constraints \cite{deNiverville:2011it}, which facilitates comparison between \JSNS\ and LSND.}

Although no angular reconstruction cuts are possible with \JSNS, precluding the possibility of identifying separate $e^+e^-$ tracks, we still have the capability of optimizing the signal window in total lepton energy $E_e$ to maximize $S/\delta B$. The left panel of Fig.~\ref{fig:OptimalCuts} shows our choice of signal windows for two representative mass splittings $\Delta = 0.1 m_1$ and $\Delta = 0.4 m_1$, in the dark photon model. To be conservative, we choose a constant bin width of 10 MeV, although we note that at energies near the 20 MeV threshold, the detector resolution is $\Delta E \lesssim 1\ \mathrm{MeV}$.

In the right panel of Fig.~\ref{fig:OptimalCuts} we show spectra of $E_e$ from DM events for $\Delta = (0.1, 0.4)m_1$ with $m_1 = 100 \ \MeV$. When decays dominate, as they do for this choice of $m_1$, the signal rate is peaked near $E_e\sim \Delta$, explaining the shape of the optimal cut curves. Indeed, in the long-lifetime regime, the probability of decay inside the detector is inversely proportional to the boost factor $\gamma$ [see Eq.~(\ref{eq:probs})], and the visible energy is $E_e \sim \gamma \Delta$, so most of the decay events have $\gamma = \mathcal{O}(1)$ and $E_e \sim \Delta$.  For most of the parameter space we consider here, $\Delta \lesssim 20\ \mathrm{MeV}$ and the maximum of the signal spectrum is below the assumed detector energy threshold, so that the lowest accessible energy bin is preferred. Because the signal rate tends to fall off much more sharply than the backgrounds away from its maximum, we find that choosing the energy bin that maximizes $S$ is a reasonable approximation to the maximum $S/\delta B$, regardless of the shape of the background.


\subsection{Dark photon model}

In Fig.~\ref{fig:DarkPhotonReachDelta} we show the reach of \JSNS\ to the dark photon model with mass splittings $\Delta = 0.1 m_1$ and $0.4 m_1$, alongside other constraints from LSND and E137~\cite{Bjorken:1988as,Batell:2014mga}, and projections from other beam dump experiments. Our results are presented in terms of the dimensionless variable $y$ (see Refs.~\cite{Izaguirre:2015yja,Krnjaic:2015mbs} for a discussion) which
is proportional to the coannihilation cross section for $s$-channel $\apr$ exchange
\be
\hspace{-0.1cm}\sigma v(\chi_1 \chi_2  \to \bar ff )      \propto \frac{y}{ m_{1}^2}      ~~,~~          y \equiv \epsilon^2 \alpha_D \left( \frac{m_1}{m_{\apr}}\right)^4~.
\label{eq:sigmavapprox}
\ee
For clarity, in Eq.~(\ref{eq:sigmavapprox}) we have approximated $m_{\apr} \gg m_{1,2}$ and dropped terms of order $\Delta/m_{\apr}$  which are always small for the benchmark parameters shown in Fig.~\ref{fig:DarkPhotonReachDelta}; in the numerical simulations, these 
approximations are not made. For each value of $m_1$, there is a critical value of $y$ for which solutions to Eq.~(\ref{eq:boltzmann}) yield the correct relic density of $\chi_1$, $\Omega_1 = \Omega_{\rm DM}$, in the present-day universe. Since
$y$ is defined as a product of independent model parameters, this critical value is insensitive to their ratios, which reduces the dimensionality of the parameter space. 
However, even though the annihilation cross section itself is insensitive to $\Delta$, large values of this parameter deplete more of the $\chi_2$ due to their decays
prior to  freeze-out, so to compensate for this depletion of coannihilation partners, larger cross sections are needed to achieve the observed relic density (see Ref.~\cite{Izaguirre:2017bqb} for 
a discussion of this early universe cosmology). This feature explains why the parameter space for $\Delta = 0.4 m_1$ requires larger $y$ values to 
achieve thermal freeze out through $\chi_1 \chi_2$ coannihilation. 
 
As anticipated, the reach for \JSNS\ is weaker than LSND for $m_1 + m_2 < m_\pi$ due to the larger neutrino backgrounds, most notably from kaons. This highlights an interesting feature of proton beam dump experiments, where for an elastic scattering signal in the absence of kinematic thresholds, the advantages afforded by going to higher beam energies are somewhat outweighed by the larger neutrino backgrounds. Still, \JSNS\ can probe unexplored parameter space between the $\pi^0$ and $\eta$ kinematic thresholds, with reach exceeding that of a similar projected search at MiniBooNE~\cite{Izaguirre:2017bqb}.  Consistent with the results of Ref.~\cite{Izaguirre:2017bqb}, we see that for inelastic DM, the strongest signal comes from decay inside the detector when kinematically allowed ($\Delta > 2m_e$). The upper boundaries of the red contours (visible for $\Delta = 0.4$, but mostly outside the plot range for $\Delta = 0.1$) are set by requiring that enough $\chi_2$ arrive at the detector before decaying. For $\Delta = 0.4$, $\chi_1$ upscattering off electrons is kinematically forbidden for $m_1 \gtrsim 40 \ \MeV$; while upscattering off nucleons could potentially improve the reach at large $m_1$, we conservatively neglect this channel for the reasons given in Sec.~\ref{sec:signals}.


\subsection{Dipole model}

\begin{figure}[t]
\centering
\hspace{-0.5cm}\includegraphics[width=0.42\textwidth]{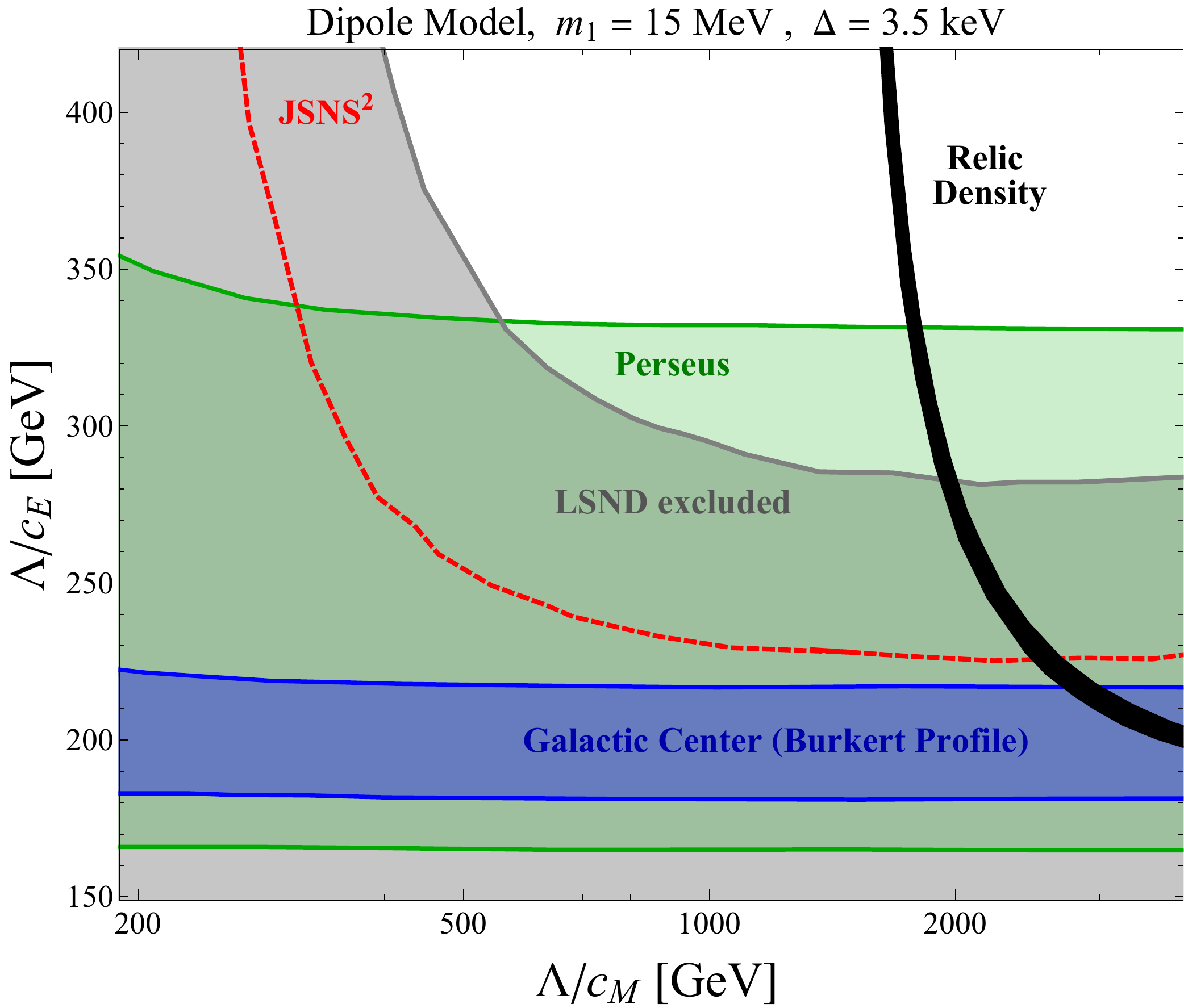}
\caption{
Parameter space for the dipole model in terms of $c_M$ and $c_E$, the electric and magnetic dipole
coefficients, respectively.  The shaded green band labeled Perseus is the region for which the inelastic dipole DM model 
can accommodate the anomalous 3.5 keV line from the Perseus Cluster, the shaded blue band is where the same model accommodates a similar excess observed from the Galactic Center assuming a Burkert halo profile, and the black curve represents the 
 parameter region for which the dipole model achieves the observed DM relic density via annihilations to SM particles;
 these regions are all taken from Ref.~\cite{DEramo:2016gqz}. 
Also shown are new constraints computed in this paper using LSND data and projections for \JSNS. 
LSND already rules out the parameter space for the Galactic Center favored region for the 3.5 keV line from Ref.~\cite{DEramo:2016gqz}. \JSNS\ can also constrain parameter space below and to the left of the red dashed curve, which also includes the favored parameter space. }
\label{fig:DipoleReach}
\end{figure}

In Fig.~\ref{fig:DipoleReach}, we show the reach of both \JSNS\ and LSND to the dipole DM model for the preferred parameter space of $m_1 = 15 \ \MeV$, $\Delta = 3.5 \ \keV$ identified in Ref.~\cite{DEramo:2016gqz}. This plot is analogous to Fig. 8 of Ref.~\cite{DEramo:2016gqz}, where the axes are the inverses of the couplings $c_E/\Lambda$ and $c_M/\Lambda$, such that regions below and to the left of the constraint curves are excluded. We see that the entire region matching the observed flux from the Galactic Center is excluded by LSND, including in particular the region which intersects the relic density curve. The projected reach of \JSNS\ is similar in shape but slightly weaker than LSND. This is driven largely by the higher neutrino backgrounds at \JSNS. Since the dipole interaction proceeds through a massless photon, the differential scattering rate increases sharply at low electron recoil energies, so a lower threshold than the 20~MeV taken for this analysis could potentially improve the \JSNS\ reach significantly. Nonetheless, \JSNS\ should also be sensitive to the preferred region for the flux from the Galactic Center.

\section{Conclusion}
\label{sec:Conc}

High-intensity neutrino experiments remain an important component of the intensity frontier program, especially in their ability to discover or falsify models of DM which are invisible at traditional direct detection experiments. We have analyzed the capability of the \JSNS\ experiment to probe two such models, where pseudo-Dirac DM has a mass splitting and interacts (and obtains its relic density) either through a dark photon or a dipole operator. Given the capabilities of the \JSNS\ neutrino detector and significant $\eta$ production at the target, the experiment can extend the existing LSND constraints on these models for DM which is too heavy to be produced from $\pi^0$ decay. In the case of the dipole DM model, we have determined that existing LSND constraints are already sufficient to rule out the preferred parameter space consistent with both thermal DM and the 3.5 keV excess from the Galactic Center, a constraint which can also be verified at \JSNS.

The reach of \JSNS\ could be significantly improved if the J-PARC MLF source were paired with a neutrino detector with the angular resolution capability to resolve $e^+e^-$ pairs resulting from the heavier DM state decaying inside the detector. This striking signal of two coincident charged tracks is likely to be essentially background-free at any neutrino experiment, and would serve as an important test of the DM models we consider here.

\section*{Acknowledgments}

YK and GK thank Eder Izaguirre and Stefano Profumo for helpful conversations, as well as the Kavli Institute for Theoretical Physics for hospitality while portions of this work were completed. Research at KITP is supported in part by the National Science Foundation under Grant No. NSF PHY-1748958. Fermilab is operated by Fermi Research Alliance, LLC, under Contract No. DE-AC02-07CH11359 with the US Department of Energy. JRJ is supported by the National Science Foundation Graduate Research Fellowship under Grant No. DGE-1256260.

\appendix

\section{Signal matrix elements, cross sections, and rates}

\label{app:Matrix}
Here we present some details on the matrix elements used to compute production and detection cross sections and rates.

\subsection{DM production}
For most of the parameter space considered in this paper, meson decay in the dark photon model yields an on-shell $A^\prime$ via $m^0 \to \gamma A^\prime$, which is allowed when $m_{A'} < m_{m^0}$. Assuming a $100\%$ branching fraction of $A'$ into $\chi_1 \chi_2$, the partial width is
\be
\Gamma_{m^0 \to \gamma \chi_1 \chi_2} = \Gamma_{m^0 \to \gamma \gamma} \times 2\epsilon^2 \left(1 - \frac{m^2_{A'}}{m^2_{m^0}}\right)^3.
\ee
For the more general expression when the decay proceeds through an off-shell $A'$, see Ref.~\cite{Kahn:2014sra}. 


For the dipole model, the mass splitting is $\Delta = 3.5 \ \keV \ll m_{\chi_1, \chi_2}$ so we can safely work in the degenerate limit $m_{\chi_1} = m_{\chi_2} = m_\chi$. The decay matrix element is
\bea
\langle |\mathcal{A}^{(E,M)}_{\pi^0 \to \gamma \chi \overline{\chi}}|^2\rangle  &=& \frac{\alpha^2 c_{E, M}^2 }{4\pi^2 \Lambda^2 f_{m^0}^2}\frac{1}{q^2}  \bigl[  (q^2-m_{m^0}^2)^2 (q^2 \pm 2m_\chi^2) \nonumber \\
 && ~~~~~~~~~~~  - 8 q^2 (k_1 \cdot k_2)(k_1 \cdot k_3) \bigr],
\eea
where the upper (lower) sign corresponds to the electric (magnetic) contribution and there is no interference between electric and magnetic dipole interactions. 
The differential $m^0$ decay width is
\bea
\label{eq:GammaM0}
\hspace{-0.15 cm}\frac{  d\Gamma_{\mzero\to\gamma\chi_1\chi_2} }{d\Omega_\gamma^* d\Omega_{\chi}^*dq^2}
\! = \!  \frac{       \langle\left|\mathcal{A}_{\mzero\to\gamma\chi_1\chi_2}\right|^2\rangle    }{4096 \pi^5  m_{\mzero}}  
 \beta(m_{\mzero}^2,0,q^2)  \beta(q^2,m_1^2,m_2^2), \nonumber \\ 
\eea
where $d\Omega_{\gamma}^*$ refers to angles in the $\mzero$ rest frame, $d\Omega_{\chi}^*$ refers to angles in the $\chi_1\chi_2$ CM frame, and
\be
\label{eq:beta}
\beta(a,b,c) = \sqrt{1 - \frac{2(b        +c )}{a} + \frac{(b-c)^2}{a^2}}.
\ee
The physical kinematic regime is $(m_1 + m_2)^2 \leq q^2 \leq m_{m^0}^2$.

\subsection{DM-electron scattering}
For the dark photon model, the process $\chi_1(k_1) e^-(p_1) \to \chi_2(k_2) e^-(p_2)$ proceeds through a $t$-channel $A'$. Neglecting terms quadratic in $m_e$, the spin-averaged squared matrix element is 
\bea
\hspace{-1cm} \langle |{\cal A}^{(A')}_{\chi_1 e \to \chi_2 e}|^2 \rangle &=&  
\frac{32 \pi^2 \epsilon^2 \alpha \alpha_D  }{(t - m^2_{A^\prime})^2} \biggl[   2(s- m_\chi^2 )^2  + 2 st + t^2     \nonumber \\ 
&& ~~~~~~~~~~~~~~~~~~~~~~~ - 4  m_\chi \Delta (s-m_\chi^2 )   \biggr],~~
 \eea
where  $s = (p_1 + k_1)^2$ and $t = (p_2 - p_1)^2$ are the usual Mandelstam variables. Here we have defined $m_\chi \equiv m_1$ and  $\Delta \equiv m_2 - m_1$ and 
neglected terms of order $(\Delta/m_A^\prime)^2\ll 1$, which are small corrections for the parameter space we consider in this paper. 

For the dipole model, scattering proceeds through a $t$-channel photon. Since we are primarily interested in small $ \Delta \sim \keV$ scale mass splittings, we
approximate the relativistic scatter as quasi-elastic where  $m_1 = m_2 = m_\chi$ and neglect terms quadratic in $m_e$. The squared amplitude in this regime is 
\bea
 \hspace{-1cm} \langle |\mathcal{A}^{({\rm dip.})}_{\chi e \to \chi e}|^2\rangle &=& -  \frac{16 \pi \alpha }{\Lambda^2 t } \biggl[    (c_E^2 + c_M^2 )(s-m_\chi^2)^2  \nonumber \\ &&
  ~~~~~~~~~~~~~~+ c_E^2 s t + c_M^2(s-m_\chi^2)t  \biggr],~~~
\eea
where we note that  $t < 0$ in the physical region.

For both models, the differential scattering cross section in the lab frame is  
\be
\frac{d\sigma}{dE_{R}} = \frac{  m_{e} \langle |{ \cal A}_{\chi_1 e \to \chi_2 e}|^2 \rangle}{  32 \pi  s  \left|  {\vec p}^*  \right|^{2} },
\ee
where $E_R$ is the energy of the recoiling electron in the lab frame, $s=(k_1 + p_1)^2 = m_1^2 + m_e^2 + 2m_e E_{\chi_1}$, and
\be
  |\vec p^{\,*}|^2 =  \frac{ (s-m_e^2-m_1^2)^2 - 4 m_e^2 m_1^2}{4s},
\ee
is the initial state three-momentum in the CM frame.

\subsection{DM decay}
In the dipole model, the heavier DM state can decay via $\chi_2 \to \chi_1 \gamma$ whose rest frame width is
\be
\Gamma(\chi_2 \to \chi_1 \gamma) = \frac{(c_E^2 + c_M^2) \Delta^3}{\pi \Lambda^2},
\ee
which is valid in the the $\Delta \ll \Lambda, m_{1,2}$ limit appropriate for our analysis. For the $\Delta \sim \keV$ splittings of interest in this paper, the $\chi_2$ is
 too long lived for an appreciable detector signature as it traverses the downstream detector, so we focus on scattering  processes instead.

In the dark photon model, decay signatures are important in the  $\Delta > 2m_e$ regime in which the $\chi_2 \to \chi_1 e^+ e^-$ process dominates the experimental reach at \JSNS. 
As mentioned in the text, for $\Delta < 2m_e$, the only decay channel available is $\chi_2 \to \chi_1 + 3 \gamma$, which is further loop-suppressed and renders $\chi_2$  stable on the length scales appropriate for beam dump experiments, so we emphasize the $\Delta > 2m_e$ regime for the remainder of this analysis. 

The squared matrix element for $\chi_2(p_1)\to\chi_1(k_1)e^+(k_2)e^-(k_3)$ decay in the dark photon model can be written
\begin{align}
&\langle\left|\mathcal{A}_{\chi_2\to\chi_1e^+ e^-}\right|^2\rangle = \frac{16\epsilon^2 e^2g_D^2}{(q^2-m_{A'}^2)^2+m_{A'}^2\Gamma_{A'}^2} \nonumber \\
&\times \left[(k_2\cdot k_1)(k_3\cdot p_1)
  + (k_2\cdot p_1)(k_3\cdot k_1) \right.\nonumber \\
 & \left.  + m_e^2(k_1\cdot p_1)
  -m_1m_2(k_2\cdot k_3)
  - 2m_1m_2m_e^2\right],
\end{align}
where $q\equiv p_1-k_1$. As in previous studies~\cite{Izaguirre:2017bqb} we only consider $m_{A'}>m_1$ and $\Delta<m_1$, so the $A'$ is always off-shell and we never make the narrow width approximation for $\chi_2$ decays. The differential decay width is 
\bea
\label{eq:GammaChi2}
\frac{   d\Gamma_{\chi_2}  }{         d\Omega_1^* d\Omega_e^*dq^2      }  \!
          =\!  \frac{\langle\left|\mathcal{A}_{\chi_2\to\chi_1e^+ e^-}\right|^2\rangle}{4096 \pi^5 m_2}    \beta(m_2^2,m_1^2,q^2)  \beta(q^2,m_e^2,m_e^2) , \nonumber \\  
\eea
where $\beta$ is defined as in \Eq{eq:beta}, $d\Omega_1^*$ refers to angles in the $\chi_1$ rest frame, and $d\Omega_e^*$ refers to angles in the $e^+e^-$ CM frame. Here the physical kinematic regime is $4 m_e^2 \leq q^2 \leq \Delta^2$. In the $\Delta \ll m_i, m_{A^\prime}$ limit with $m_e \to 0$, the total $\chi_2$ width has a simple closed form
\be
\Gamma(\chi_2 \to \chi_1 e^+e^-) = \frac{4 \epsilon^2 \alpha \alpha_D \Delta^5}{15 \pi m_{A^\prime}^4},
\ee
up to corrections of order $(\Delta/m_{A^\prime})^6$.

Given the partial width of mesons $m^0$ into DM from integrating Eq.~(\ref{eq:GammaM0}) over 3-body phase space, the flux of excited state DM is
\be
N_{\chi_2} = \sum_{m^0} N_{m^0} \frac{\Gamma_{m^0 \to \gamma \chi_1 \chi_2}}{\Gamma_{m^0}},
\ee
where $N_{m^0}$ is the flux of each species of meson which can decay into DM.  For each $\chi_2$, the probability of decay inside the detector is 
\be \label{eq:probs}
 P_{\rm survive} \,   P_{\rm decay} = e^{-\Gamma_{\chi_2} d/\beta\gamma} \left(1 - e^{-\Gamma_{\chi_2} \ell/\beta\gamma} \right),
 \ee
which is the product of the probability for $\chi_2$ to survive to a distance $d$ from the beam dump and the probability to decay after traversing a path length $\ell$ inside the detector. Here, $\gamma = E_{\chi_2}/m_2$ is the boost factor, $\beta = v/c$ is the $\chi_2$ velocity, and $\Gamma_{\chi_2}$ is obtained by integrating Eq.~(\ref{eq:GammaChi2}) over 3-body phase space. We compute the total number of decay events using a Monte Carlo simulation normalized to the total flux $N_{\chi_2}$; see Ref.~\cite{Izaguirre:2017bqb} for further details.

\bibliography{JSNS2DMBib.bib}

\end{document}